\documentclass[aps,prl,showpacs,twocolumn,,superscriptaddress]{revtex4-1}

\usepackage{amsmath}
\usepackage{amsfonts}
\usepackage{amssymb}
\usepackage{epsfig}

\usepackage{graphicx}

\usepackage{hyphenat}
\usepackage{hyperref}
\usepackage{placeins}	
\usepackage{wasysym}
\usepackage{color}

\usepackage{array}
\newcolumntype{C}[1]{>{\centering\arraybackslash}m{#1}}
\usepackage{overpic}

\begin{document}

\title{Spin liquid and quantum phase transition without symmetry breaking in a frustrated three-dimensional Ising model}

\author{Julia R\"ochner}
\affiliation{Lehrstuhl f\"ur Theoretische Physik 1, TU Dortmund, Germany}
\email{julia.roechner@tu-dortmund.de}

\author{Leon Balents}
\affiliation{Kavli Institute for Theoretical Physics, University of California, Santa Barbara, CA 93106-4030}
\email{balents@physics.ucsb.edu}

\author{Kai Phillip Schmidt}
\affiliation{Lehrstuhl f\"ur Theoretische Physik I, Staudtstra{\ss}e 7, Universit\"at Erlangen-N\"urnberg, D-91058 Erlangen, Germany}
\email{kai.phillip.schmidt@fau.de}

\begin{abstract}
We show that the highly frustrated transverse-field Ising model on the three-dimensional pyrochlore lattice realizes a first-order phase transition without symmetry breaking between the low-field Coulomb quantum spin liquid and the high-field polarized phase. The quantum phase transition is located quantitively by comparing low- and high-field series expansions. Furthermore, the intriguing properties of the elementary excitations in the polarized phase are investigated.   We argue that this model can be achieved experimentally by applying mechanical strain to a classical spin ice material comprised of non-Kramers spins such as  Ho$_2$Ti$_2$O$_7$.  Taken together with our results, this provides a new experimental platform to study quantum spin liquid physics.
\end{abstract}

\maketitle


The appearance of new collective degrees of freedom in strongly correlated systems is a pervasive theme in physics, exemplified in magnetism by spin liquids -- highly correlated states of spins with novel excitations and emergent gauge structures \cite{Balents2010,Savary2016}.  Amongst the most storied examples of the latter are the spin ice pyrochlores, Ho$_2$Ti$_2$O$_7$ or Dy$_2$Ti$_2$O$_7$ where strong geometric frustration among coupled magnetic moments gives rise to classical spin liquids with defects that behave like magnetic monopoles \cite{Castelnovo2008,Fennel2009,Jaubert2009}.

Quantum effects are essentially zero in these spin-ice compounds, which are described accurately by classical Ising models -- i.e.~only the (local) $\sigma_i^z$ component of the spins appears in the Hamiltonian.  However, theoretically, a quantum version of spin ice is highly desirable. One expects the presence of a so-called Coulomb quantum spin liquid (CQSL) \cite{Hermele2004,Shannon2012,Balents2010} with gapped electric and magnetic excitations as well as an emergent photon.  Quantum fluctuations may be introduced by additional exchange interactions involving spin flips (e.g.~XY or more complex couplings), which naturally occur in some other pyrochlores  like Yb$_2$Ti$_2$O$_7$ \cite{Ross2011,Pan2016}.  However, such quantum exchange models are quite complex, and their phase diagrams contain many other ground states in addition to the desired CQSL \cite{PhysRevLett.108.037202,PhysRevB.86.104412}, so achieving the right type of quantum exchange requires some serendipity.

\begin{figure} [t!]
 \includegraphics[width=\columnwidth]{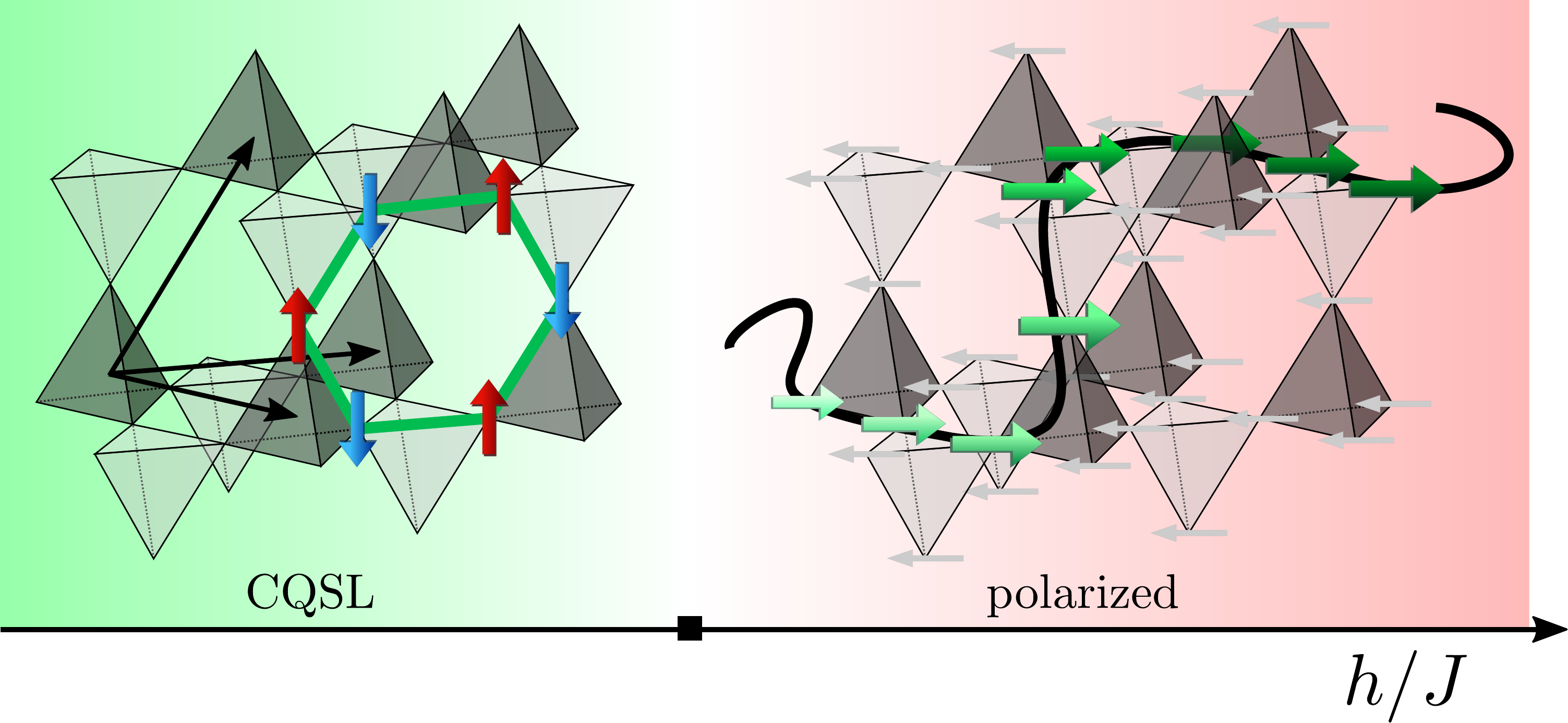}
 \caption{Phase diagram of the TFIM on the pyrochlore lattice as a function of $h/J$ consisting of the low-field CQSL and the high-field polarized phase separated by a first-order phase transition at $h_{\rm c}/J_{\rm c} \approx 0.6$ shown as a black filled square. Four-site unit cells of the pyrochlore lattice are shown as dark tetrahedra. {\it Left illustration}: Dark arrows denote the three unit cell vectors. Green  hexagon exemplifies a hexagon on which the ring-exchange $K$-term in Eq.~\eqref{eq:H_eff_lf} acts on alternating spins. {\it Right illustration}: Moving elementary spin-flip excitation above the high-field polarized state.}
 \label{fig:pyrochlore}
\end{figure}

At the model Hamiltonian level, a simpler route to ``quantum-ize'' classical spin ice is to add a transverse field.  This is not achievable with a physical magnetic field, however, because the latter couples most strongly to the Ising spin components, i.e. it introduces longitudinal fields which quench fluctuations instead of enhancing them.  However, it has been recently pointed out that for non-Kramer's rare earth ions like Ho$^{3+}$ or Pr$^{3+}$, local electric field gradients play the role of transverse fields in the spin Hamiltonian while preserving time-reversal symmetry \cite{Savary2016}.  This provides a mechanism to induce transverse fields while protecting the system from longitudinal fields.  In Ref.~\onlinecite{Savary2016}, it was suggested to use disorder to induce random transverse fields.  Here, we consider a simpler possibility: {\em straining} a non-Kramers spin ice material to lower the local symmetry and thereby create a {\em uniform} transverse field.  Large strains are achievable in thin films, which for spin ice materials have recently been grown \cite{bovo2014restoration}. The low-energy physics of such systems is expected to be described to a good extent by the transverse-field Ising model (TFIM) on the three-dimensional pyrochlore lattice.

The TFIM is one of the archetypal models used in various areas in physics and is known to host a plethora of interesting physical
 phenomena, especially on highly frustrated lattices \cite{Moessner2001,Moessner2000,Powalski2013,Sikkenk2016}. At the same time the theoretical treatment of three-dimensional frustrated systems including TFIMs represents a notable challenge and quantitative results are hard to extract. In this work we apply low- and high-field series expansions to determine the phase diagram of the three-dimensional pyrochlore TFIM quantitatively as shown in Fig.~\ref{fig:pyrochlore}. We find a first-order quantum phase transition without symmetry breaking separating the CQSL at low fields from the high-field polarized phase, and locate the quantum phase transition point quantitatively.  

{\it{Model:}} We study the Ising model in a transverse magnetic field on the pyrochlore lattice
\begin{equation}
\label{eq:ising}
	\mathcal{H}=J \sum_{<i,j>}{\sigma^z_i \sigma^z_j}-h\sum_i{\sigma^x_i}
\end{equation}
with Pauli matrices $\sigma_i^{\alpha}$ acting on site i, the antiferromagnetic nearest-neighbour exchange $J>0$, and the strength of the transverse magnetic field $h$. The pyrochlore lattice has a four site unit cell and we use the three unit cell vectors illustrated as black arrows in the left illustration in Fig.~\ref{fig:pyrochlore} to define the three-dimensional momentum $\vec{k}$. In the limit of large fields the system is in the polarized phase and elementary excitations are dressed spin flips (see also right illustration in Fig.~\ref{fig:pyrochlore}). The situation is drastically different in the other limit of small fields. For $h=0$, one has extensively many ground states which fulfill the so-called ``two-in-two-out'' ice rule, i.e.~if we regard $\sigma_i^z$ as the spin component along the local $\langle 111\rangle$ axis of site $i$, then any state which has two spins pointing into and two spins point out of each tetrahedron is a ground state.  In our variables, this simply corresponds to states with two up (+1) and two down (-1) spins in each tetrahedron. An infinitesimal field $h$ introduces quantum fluctuations into the classical spin liquid and lifts the extensive degeneracy. The resulting ground state, as argued in Ref.\cite{Savary2016}, is a CQSL exhibiting exotic excitations like electric and magnetic monopoles as well as an emerging photon. In the following we consider both limits of the TFIM in order to pinpoint the quantum phase transition between the CQSL and the polarized phase.

{\it{High-field limit:}} We use perturbative continuous unitary transformations (pCUTs) \cite{Knetter2000,Knetter2003} to set up the high-order series expansions in the limit $J/h\rightarrow 0$. To this end we introduce hardcore boson creation and annihilation operators $b^\dagger_i$, $b^{\phantom{\dagger}}_i$, and \mbox{$\hat{n}_i\equiv b^\dagger_i b^{\phantom{\dagger}}_i$} on site $i$ using the Matsubara-Matsuda transformation \cite{Matsubara1956} in the x-basis $\sigma_i^z=b^\dagger_i+b^{\phantom{\dagger}}_i$ and $\sigma_i^x=1-2\hat{n}_i$. This allows to rewrite Eq.~\eqref{eq:ising} as
\begin{align}
\label{eq:H_tfim_orig_boson}
  \frac{\mathcal{H}}{2h} &=-\frac{N}{2}+\sum_{j} \hat{n}_j + x \sum_{<i,j>} \left( b^\dagger_i b^\dagger_j + b^\dagger_i b^{\phantom{\dagger}}_j + {\rm H.c.}\right) \nonumber\\
                         &=-\frac{N}{2}+\hat{Q}+x\left( \hat{T}_{+2}+\hat{T}_{0}+\hat{T}_{-2}\right)\, ,
\end{align}
where $N$ is the number of sites, $x\equiv J/2h$ is the expansion parameter in the high-field limit, and $\hat{Q}$ counts the number of hardcore bosons $q$. The unperturbed Hamiltonian $\mathcal{H}_0\equiv -\frac{N}{2}+\hat{Q}$ has therefore an equidistant ladder spectrum bounded from below. Each elementary energy quantum signals the presence of a hardcore boson which corresponds to a spin flip above the fully polarized spin state in the original spin language. The perturbation $\mathcal{V}\equiv x\sum_n \hat{T}_n$ is written in terms of operators $\hat{T}_n$ which change the number of hardcore bosons by $n$. The pCUT method maps Eq.~\eqref{eq:H_tfim_orig_boson}, order by order in $x$, to an effective Hamiltonian $\mathcal{H}_{\rm eff}$ which commutes with $\hat{Q}$, i.e.~the effective Hamiltonian is block-diagonal and conserves the number of quasi-particles ($qp$). We have therefore mapped the complicated many-body problem to a few-body problem so that each $qp$-block can be tackled separately. 

\begin{figure} [t!]
 \includegraphics[width=\columnwidth]{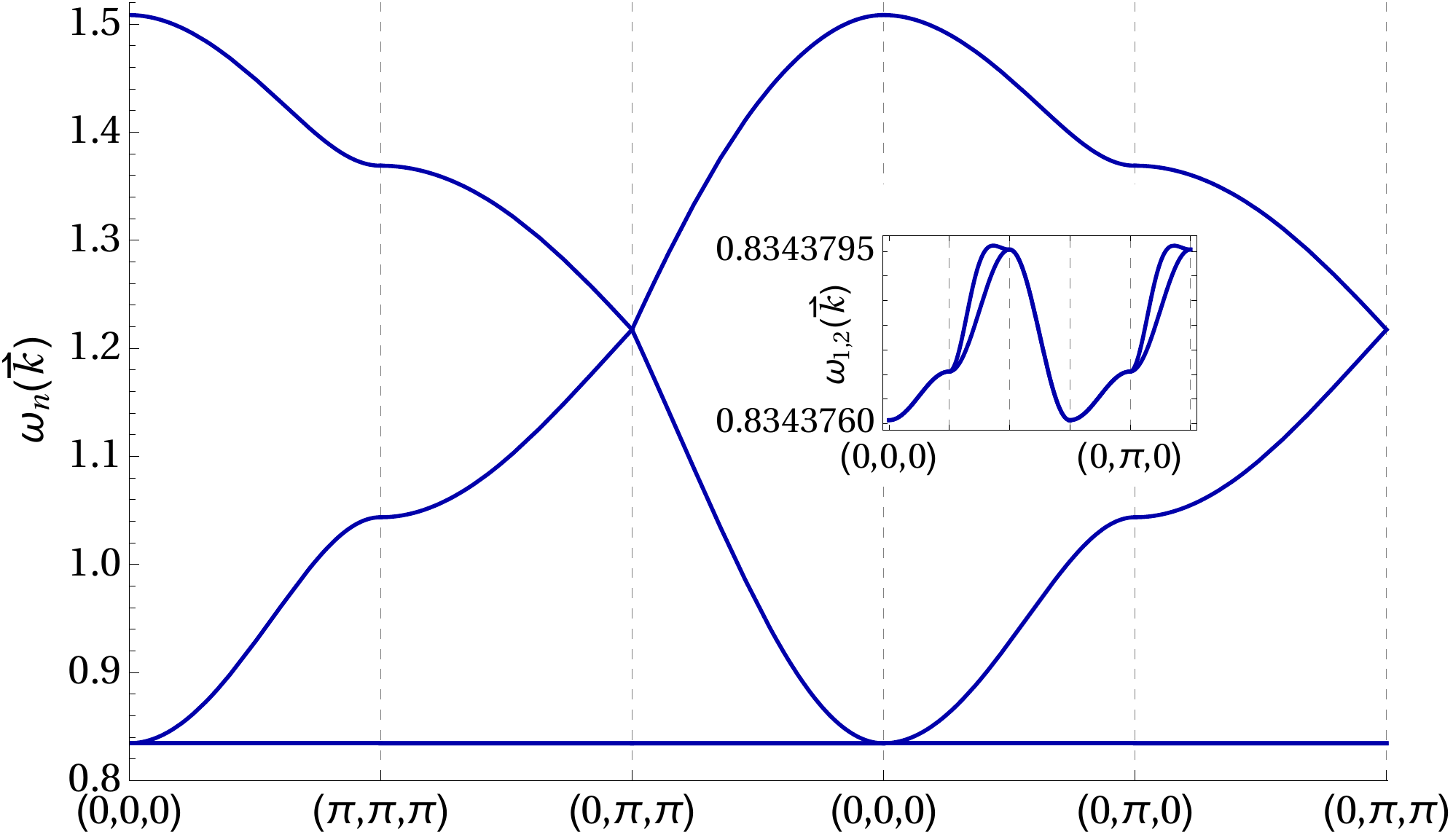}
  \caption{Illustration of the four $1qp$ bands $\omega_n (\vec{k})$ with $n\in\{1,2,3,4\}$ in the polarized phase for $x=0.1$ along a characteristic path in the three-dimensional Brillioun zone using the bare order-$11$ series expansion. The lowest band is two-fold degenerate for all momenta except for $\vec{k}=(0,0,0)$ where the degeneracy is three. {\it Inset}: Zoom on the two low-energy bands.}
 \label{fig:hf_disp}
\end{figure}

In this work we have focused on the $0qp$ and $1qp$ channel. The effective Hamiltonian in the $0qp$ channel is just the ground-state energy per site $\epsilon_0^{\rm hf}$. In the $1qp$ channel one gets an effective one-particle hopping Hamiltonian
\begin{align}
\label{eq:H_eff_1qp}
  \frac{\mathcal{H}_{\rm eff}^{1qp}}{2h} &=\sum_{i}\sum_{\delta} a_\delta \left( b^\dagger_i b^{\phantom{\dagger}}_{i+\delta} + {\rm H.c.}\right)\, ,
\end{align}
where the $a_\delta$ are the one-particle hopping amplitudes. We have calculated $\epsilon_0^{\rm hf}$ and the $a_\delta$ as high-order series expansions up to order 11 in $x$. This is achieved by exploiting the linked-cluster theorem and a full-graph decomposition. In the $1qp$ sector one has 1056 topologically distinct graphs. The resulting ground-state energy per site reads
\begin{align}
\label{eq:e0_hf}
  \frac{\epsilon_0^{\rm hf}}{2h} =&-\frac{1}{2} - \frac{3}{2}x^2 + 3 x^3 - \frac{57}{8}x^4+ \frac{93}{4} x^5- \frac{867}{8} x^6\nonumber\\ 
                               & + \frac{5235}{8} x^7- \frac{589953}{128} x^8 + \frac{8660373}{256} x^9\nonumber\\ 
                               & - \frac{252903465}{1024}x^{10}+ \frac{3696508953}{2048} x^{11}\,. 
\end{align}
The $1qp$ channel can be simplified by applying a Fourier transformation to momentum space. This results in the four $1qp$ bands $\omega_n(\vec{k})$ due to the four-site unit cell of the pyrochlore lattice. A representative plot for these bands is shown in Fig.~\ref{fig:hf_disp} for $x=0.1$ using the bare order-$11$ series. 

\begin{figure} [t!]
 \includegraphics[width=\columnwidth]{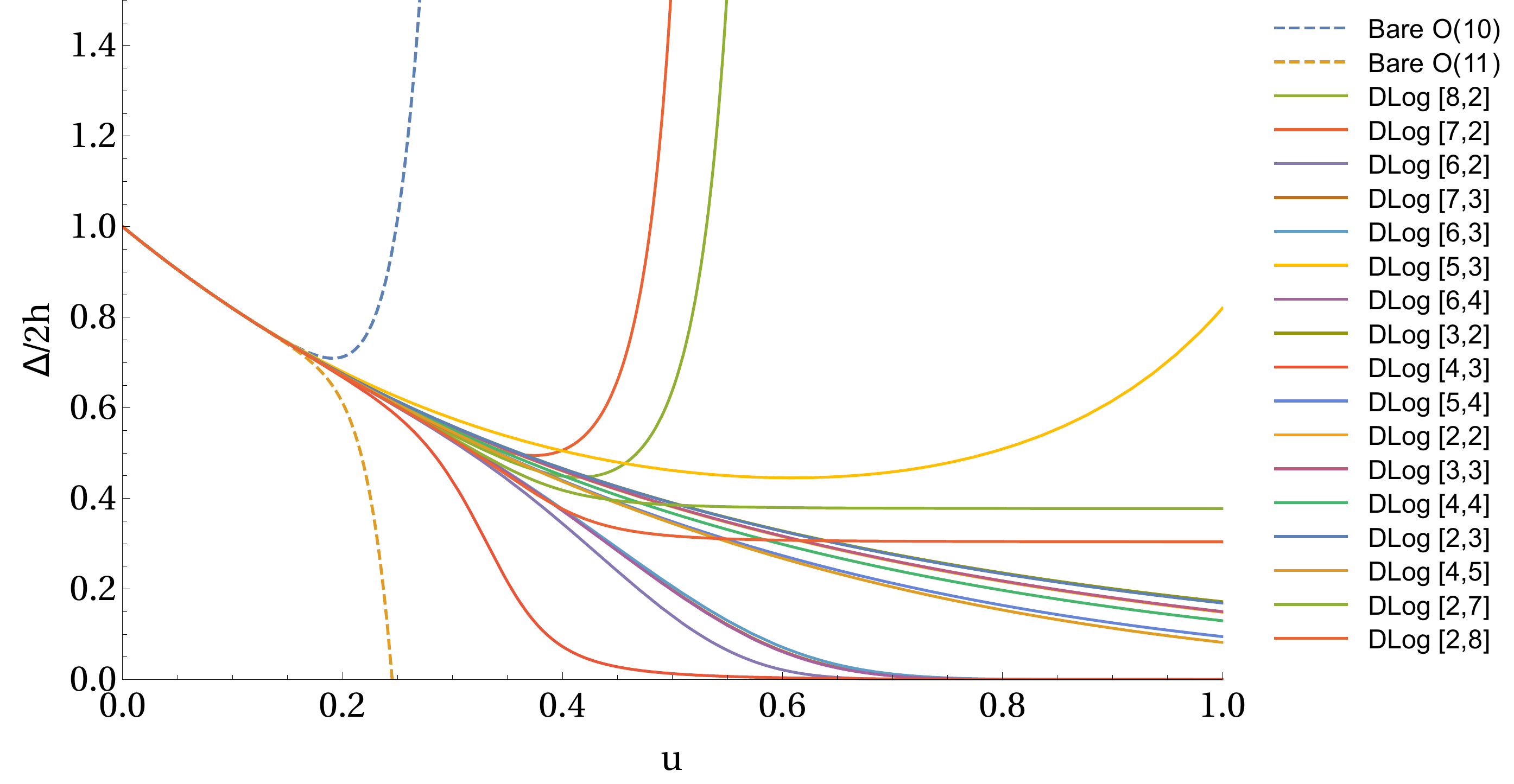}
 \caption{High-field gap $\Delta/h$ as a function of \mbox{$u=x/(x+1)$} using the bare order-$11$ series of the high-field expansion. The bare series of order $10$ and $11$ are shown as dashed lines. Other lines refer to various DlogPad\'{e} extrapolants $[n,m]$.} \label{fig:hf_gap}
\end{figure}

The $1qp$ band structure has several interesting features. There are two dispersive high-energy bands while the two low-energy bands are almost flat (see inset of Fig.~\ref{fig:hf_disp}). In fact, the first process which selects the specific gap momentum $\vec{k}=(0,0,0)$  occur only at order 8 in perturbation theory, similar to the TFIM on the kagome lattice \cite{Powalski2013}. At $\vec{k}=(0,0,0)$, one has an exact three-fold degeneracy of the gap $\Delta^{\rm hf}$ and the series of the gap can be extracted analytically
\begin{align}
\label{eq:gap_hf}
  \frac{\Delta^{\rm hf}}{2h} =& 1 - 2 x + 4 x^2 - 6 x^3 + 6 x^4 - 25 x^5 + \frac{901}{4} x^6\nonumber\\
                            &  - \frac{82783}{32} x^7 + \frac{458339}{16} x^8 - \frac{17244199}{64} x^9\nonumber\\ 
                               &  + \frac{872369819}{384} x^{10} - \frac{1361632112501}{73728} x^{11}\,. 
\end{align}   
In the following we would like to check whether the quantum phase transition between the polarized phase and the CQSL is second order. In this case it can be located and characterized by the closing of the gap $\Delta^{\rm hf}$. We therefore analyze the bare series as well as DlogPad\'{e} extrapolants ${\rm DLog} [n,m]$ with $n+m\leq 10$ of $\Delta^{\rm hf}$ as a function of $u=x/(x+1)$ with \mbox{$u\in[ 0,1] $} which are shown in Fig.~\ref{fig:hf_gap}. DlogPad\'{e} is a tool to extrapolate high-order series expansions. The key idea is to consider the logarithmic derivative $\partial_x \ln \Delta^{\rm hf}$ and to build standard Pad\'{e} extrapolants $[n,m]$ where $n$ ($m$) is the polynomial degree of the numerator (denominator) of the rational function $[n,m]$. In contrast to Pad\'{e} extrapolation, DlogPad\'{e}s are especially useful for quantum critical behavior, since they behave as $(x-x_{\rm c})^{\alpha}$ close to poles $x_{\rm c}$ of ${\rm DLog} [n,m]$. For an extensive review we refer to Ref.~\onlinecite{Guttmann89}. 

We find that the bare series is alternating and converges only up to $u\approx 0.2$. As a consequence, extrapolations are indeed essential to investigate the full parameter axis. Interestingly, there are no tendencies for a gap closing in all extrapolants. We stress that even the extrapolants in Fig.~\ref{fig:hf_gap}, which approach zero, do not possess any poles. Overall, no indications for a second-order quantum phase transition can be detected. However, a CQSL is expected to be present for small fields and therefore at least one phase transition must exist in the pyrochlore TFIM, which is different to the disorder by disorder scenario in the two-dimensional kagome TFIM \cite{Moessner2001,Powalski2013}. The analysis of the gap in the polarized phase therefore suggests that the quantum phase transition in the pyrochlore TFIM is first order.     

{\it{Low-field limit:}} In order to confirm and to locate the first-order quantum phase transition we will compare the ground-state energies of the CQSL and the polarized phase. 

The unperturbed Ising model has an extensive number of ground states with energy per site $-J$, since each spin is part of two tetrahedra and a tetrahedron fulfilling the two-in-two-out rule, i.e.~\mbox{$|\sigma_{\rm tot}^{z}|=|\sigma_1^z+\sigma_2^z+\sigma_3^z+\sigma_4^z|=0$} for the four spins of this tetrahedron, exhibits an energy $-2J$. Excited states correspond to spin configurations containing tetrahedra with $|\sigma_{\rm tot}^{z}|=2$ or $|\sigma_{\rm tot}^{z}|=4$, e.g.~a single spin flip on a spin-ice ground state yields always two tetrahedra with $|\sigma_{\rm tot}^{z}|=2$ attached to this spin and thus enhances the energy by $4J$. 

The low-field expansion is performed as follows. We use degenerate perturbation theory in $h/J$ to derive an effective Hamiltonian $\mathcal{H}_{\rm eff}^{\rm lf}$ which acts within the extensive ground-state manifold of spin-ice states. It is found that only even orders contribute to $\mathcal{H}_{\rm eff}^{\rm lf}$ and that the degeneracy is not lifted up to order 4. The first non-trivial process occurs in order 6. Physically, this process couples two spin-ice states which are connected by a simultaneous flip of the six spins on a single elementary hexagon pointing alternatingly up and down (see also the left illustration in Fig.~\ref{fig:pyrochlore}). Up to order 8 no other processes appear in the effective Hamiltonian and one obtains
\begin{align}
\label{eq:H_eff_lf}
  \mathcal{H}_{\rm eff}^{\rm lf}  &=e_0 N + K\sum_{\hexagon}\left( \sigma^+_i \sigma^-_j \sigma^+_k \sigma^-_l \sigma^+_m \sigma^-_n + {\rm H.c.}\right)\, ,
\end{align}
where 
\begin{align}
e_0=&-J-\frac{1}{4}\frac{h^2}{J}-\frac{7}{192}\frac{h^4}{J^3}-\frac{893}{34560}\frac{h^6}{J^5}-\frac{209966173}{6967296000}\frac{h^8}{J^7}\nonumber\\
K=& -\frac{63}{256}\frac{h^6}{J^5}-\frac{33833}{165888}\frac{h^8}{J^7}\, .
\end{align}
The leading order-6 contribution of $K$ has been already derived in Ref.~\onlinecite{Savary2016}. The sum is taken over all hexagons of the pyrochlore lattice. This model is known as the ring exchange model which is believed to realize a CQSL. In Ref.~\onlinecite{Shannon2012}, the ring-exchange model has been studied numerically by quantum Monte Carlo simulations and a ground-state energy per site \mbox{$e_0^{\rm ring}=0.189078 J$} is found. This allows us to obtain the order-$8$ low-field expansion of the pyrochlore TFIM for the ground-state energy per site $\epsilon_0^{\rm lf} \equiv e_0+K e_0^{\rm ring}$.

\begin{figure} [t!]
 \includegraphics[width=\columnwidth]{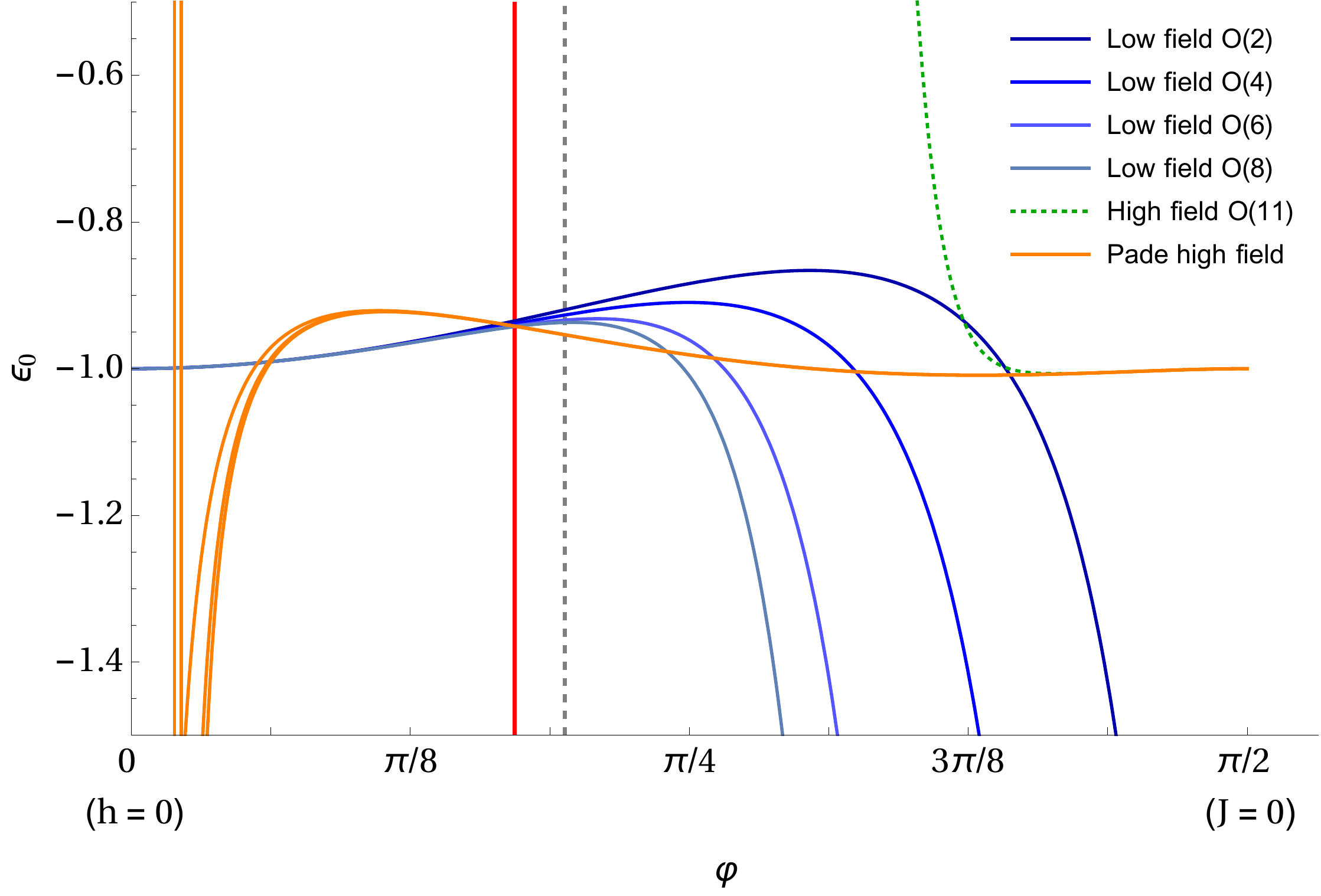}
 \caption{Comparison of the ground-state energies $\epsilon_0$ from the low- and high-field expansions as a function of $\varphi$ setting $J=\cos\varphi$ and $h=\sin\varphi$. Dashed vertical line refers to the mean-field result from Ref.~\onlinecite{Savary2016} and solid vertical line corresponds to location of the first-order quantum phase transition as obtained from the crossing of the series expansions.}
 \label{fig:e0}
\end{figure}

We are therefore in the position to compare the energies of the CQSL and the polarized phase directly.  We set $J=\cos\varphi$ and $h=\sin\varphi$ which allows a comparison between $e_0^{\rm lf}$ and $e_0^{\rm hf}$ on the full parameter axis as shown in Fig.\ref{fig:e0}. For the high-field expansion Pad\'{e} extrapolation is used which works extremely well for $e_0^{\rm hf}$, i.e.~various extrapolants lie almost on top of each other in a wide range of $\varphi$. The series of $e_0^{\rm lf}$ is monotonous and already the bare series is sufficient, since the crossing between both energies is at rather small values of $\varphi$. Using the bare order-$8$ of the low-field expansion, this crossing at $\varphi_{\rm c}\approx 0.542$ ($h/J\approx 0.602$) corresponds to the location of the first-order quantum phase transition between the CQSL and the polarized phase, which we can therefore determine quantitatively. We remark that the mean-field value $\varphi_{\rm c}^{\rm mf}\approx 0.611$ from Ref.~\onlinecite{Savary2016} is slightly larger than $\varphi_{\rm c}$, which is consistent with the expectation that the mean-field calculation overestimates the CQSL.  

{\it{Conclusion:}} We have determined quantitatively the location of the quantum critical point between the CQSL and the polarized phase in the transverse-field Ising model on the three-dimensional highly frustrated pyrochlore lattice, which is achieved via high-order series expansions in both phases. The quantum phase transition is first order and no symmetry breaking occurs on both sides of the transition.  The first order nature is actually to be expected based on field theoretic renormalization group arguments \cite{Savary2016}.  In this framework, the field-induced transition from the CQSL is a {\em confinement} transition, induced by the condensation of an elementary bosonic monopole (it can be regarded as electric or magnetic, depending upon convention).  Because no symmetry breaking occurs in the transition, the monopole carries no symmetry quantum numbers, only its emergent charge, and is thus described by a complex scalar field coupled to a U(1) gauge field. Such a field theory, formulated in imaginary time, is mathematically identical to the Ginzburg-Landau theory of a superconductor in four dimensions \cite{Peskin}. Renormalization group analysis shows that this transition is first order   \cite{HLM}.  By combining two series expansions, we have been able to access this first order transition quantitatively in a complicated three dimensional model, for which other computational techniques are very demanding.

Most importantly for experiments, the critical field defining the size of the CQSL is found to be relatively large, i.e. the CQSL is robust, promising good access to the fascinating properties of spin liquids in the laboratory. We expect our results to be of direct relevance for "quantum-izing" classical spin ice systems. The transverse field can be introduced experimentally in classical (non-Kramers) spin ice materials like Ho$_2$Ti$_2$O$_7$ by applying strain, which lowers the symmetry and results in an effective coupling between the two local spin states. We therefore propose strained classical (non-Kramers) spin ice Ho$_2$Ti$_2$O$_7$ as a new experimental platform for spin liquid physics.  

{\it{Acknowledgement:}}
We thank Frank Pollmann for providing us with the numerical ground-state energy of the ring exchange model on the pyrochlore lattice studied in Ref.~\onlinecite{Shannon2012}. Julia R\"ochner thanks the Artur- und Lieselotte-Dumcke-Stiftung and the TU Dortmund for financial support within the framework of her Deutschlandstipendium.  This research was supported by the  U.S. Department of Energy, Office of Science, Office of Basic Energy Sciences under Award Number DE-FG02-08ER46524.

\end{document}